\documentclass[twocolumn,showpacs,preprintnumbers,amsmath,amssymb]{revtex4}

\usepackage{graphicx}
\usepackage{dcolumn}
\usepackage{bm}

\begin{document}

\title{Single-shot Sub-Rayleigh Imaging with Sparse Detection}

\author{Wenlin Gong}
\email{gongwl@siom.ac.cn}
\author{Shensheng Han}
\affiliation{ Key Laboratory for
Quantum Optics and Center for Cold Atom Physics of CAS, Shanghai
Institute of Optics and Fine Mechanics, Chinese Academy of Sciences,
Shanghai 201800, China}

\date{\today}

\begin{abstract}
For conventional imaging, the imaging resolution limit is given by the Rayleigh criterion.
Exploiting the prior knowledge of imaging object's sparsity and fixed optical system, imaging beyond the conventional Rayleigh limit, which is backed up by numerical simulation and experiments, is achieved by illuminating the object with single-shot thermal light and detecting the object's information at the imaging plane with some sparse-array single-pixel detectors. The quality of sub-Rayleigh imaging with sparse detection is also shown to be related to the effective number of single-pixel detectors and the detection signal-to-noise ratio at the imaging plane.
\end{abstract}

\pacs{42.25.Kb, 42.30.Va, 42.30.Lr}

\maketitle

For standard conventional imaging of directly recording the object's intensity distribution with a charge coupled device (CCD) camera, both the imaging system's Rayleigh limit and the camera's pixel-resolution restrict the optical system's imaging resolution \cite{Rayleigh}. For example, the imaging resolution is mainly determined by the optical system's Rayleigh limit in remote sensing because the numerical aperture (\emph{N.A.}) of imaging lens is usually small relative to the detection distance. While in long-wavelength radiation band such as infrared and terahertz imaging, because the camera with large planar arrays is very hard to manufacture, the imaging resolution is mainly limited by the camera's pixel-resolution.

Compared with overcoming the camera's pixel-resolution to imaging resolution, many methods are invented to overcome the imaging system's Rayleigh limit at present \cite{Ash,Ramakrishna,Pendry,Huang,Hell,Hell1,Rust,Suarez,Goodman,Harris,Mallat,Hunt,Kolobov}. Exploiting the evanescent components at the object's immediate proximity, sub-Rayleigh imaging can be achieved, but this method is only applied in the near-field range \cite{Ash,Ramakrishna,Pendry}. Several microscopy techniques based on fluorescence are also introduced to improve imaging resolution. However, it requires scanning or repetitive experiments, thus which limits real-time applications \cite{Huang,Hell,Hell1,Rust,Suarez}. In addition, using additional a priori information of optical system, the imaging resolution beyond Rayleigh diffraction limit can be obtained. However, the improvement degree is limited in practice because of the influence of detection noise \cite{Goodman,Harris,Mallat,Hunt,Kolobov}.

Recently, the image's sparsity has been taken as popular a priori, which is a quite general assumption because most of the natural objects are sparse in a known basis (or under a suitable basis transform), an image can be stably reconstructed by sparse reconstruction technique even if the measurement number is less than Nyquist rate and this
technique, to some extent, is robust to noise \cite{Donoho,Donoho1,Donoho2,Candes,Candes1}. Moreover, this technique has already been applied to de-noising \cite{Candes,Candes1}, super-resolution imaging \cite{Gong,Gong1}, remote sensing \cite{Herman,Zhao}, compressive imaging \cite{Rowe,Duarte,Katz,Du,Gong2}, and magnetic resonance imaging with success \cite{Lustig}. For conventional imaging, the optical system is usually fixed or known, so the optical system's point spread function (PSF) can be also taken as a priori. Furthermore, based on the object's sparse priori property, measurements below Nyquist rate are required to exactly restore the object. Therefore, exploiting the object's sparsity assumption in a known basis and the prior knowledge of PSF, single-shot sub-Rayleigh imaging is possible even if some sparse-array single-pixel detectors are used to record the object's information at the imaging plane.
\begin{figure}[htbp]
\centering
\includegraphics[width=8.5cm]{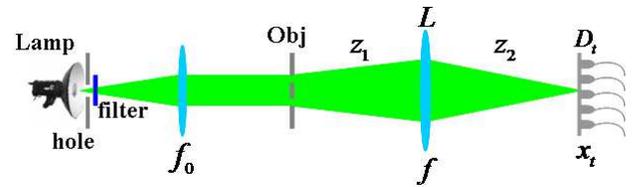}
\caption{Schematic of thermal-light single-shot imaging with sparse detection. }
\end{figure}

Fig. 1 presents experimental demonstration schematic of thermal-light single-shot sub-Rayleigh imaging with sparse detection. The uniform light emitting from a halogen lamp is filtered by an optical filter (with the center wavelength $\lambda$=650 nm and the bandwidth $\Delta\lambda$=10 nm) and then collimated by a lens with the focal length $f_0$. The object is illuminated by the collimation light and then imaged onto the imaging plane $x_t$ by a standard conventional imaging setup. Different from recording the object's transmission information by a CCD camera at the imaging plane $x_t$, the detection system $D_t$ we propose is characterized by some sparse-array single-pixel detectors.

The distances $z_1$, $z_2$ and the focal length of the lens $f$ obey Gaussian thin-lens equation: $\frac{1}{z_1}+\frac{1}{z_2}=\frac{1}{f}$. Based on Rayleigh criterion \cite{Rayleigh}, the resolution limit $\Delta x_s$ of conventional imaging shown in Fig. 1 is determined by the wavelength $\lambda$ and the \emph{N.A.} of the lens $f$, namely
\begin{eqnarray}
\Delta x_s=0.61\frac{\lambda}{N.A.}\simeq1.22\frac{\lambda z_1}{L}.
\end{eqnarray}
where $L$ is the effective transmission aperture of the imaging lens $f$ and \emph{N.A.} is approximate to $\frac{L}{2z_1}$.

When the light is fully spatially incoherent and uniform, the intensity at the imaging plane $x_t$ is
the convolution of the intensity at the object plane with the incoherent
point spread function (PSF) \cite{Goodman}
\begin{eqnarray}
I_t(x)= I_{obj}(x)\otimes h(x)+I_{noise}(x).
\end{eqnarray}
where $h(x)$ is the optical system's PSF, $I_{obj}(x)$ is the object's
intensity distribution, $I_{noise}(x)$ is the intensity distribution of the noise and $I_t(x)$ is the intensity distribution at the imaging plane $x_t$. For the optical system shown in Fig. 1,
the PSF is
\begin{eqnarray}
h(x)=\sin c^2[\frac{L}{\lambda z_1}x].
\end{eqnarray}
where $\sin c(x)=\frac{\sin(\pi x)}{\pi x}$.

From Eqs. (2) and (3), compared with the original object, it is a blurred image with low spatial resolution at the imaging
plane $x_t$ when the transmission aperture of the imaging lens $f$ is small. In order to restore
a high-resolution image, the relation shown in Eq. (2) usually yields ${\cal F}\{ I_t (x)\}  = {\cal F}\{ I_{obj} (x)\} H +{\cal F}\{ I_{obj} (x)\}$ in the spatial-frequency domain,
where ${\cal F}$ denotes Fourier transform and $H$ is the optical transfer function. Therefore, the problem is that we wish to reconstruct
the object but the measured image is smeared by a low-pass filter. Although many iterative methods were used to restore the object, the reconstruction quality strongly depends on the signal-to-noise ratio (SNR) in the measured data \cite{Goodman,Harris,Mallat,Hunt,Kolobov}. Furthermore, for the optical system shown in Fig. 1, because the detection system $D_t$ is characterized by some sparse-array single-pixel detectors and many low-frequency information will be also missed, it becomes much more difficult to restore the object by traditional iterative methods, compared with recording all the smeared image's low-frequency information by a CCD camera.

Different from the method described above, we try to directly restore the object borne on the optical system shown in Fig. 1,
exploiting the sparsity assumption of the object in a known basis and the prior knowledge of the fixed optical system. According to sparse reconstruction theory, even if the measurement process is noiseless, there are an infinite number of images, which$-$after being convoluted by the PSF$-$will result in the smeared image, our convex optimization program is how to find the sparsest one. The sparse reconstruction technique has mathematically demonstrated that if the object is sparse enough, then any sparsity-based reconstruction method is bound to find the sparsest solution \cite{Donoho,Donoho1,Donoho2,Candes,Candes1}. Here, we have employed the gradient projection for sparse reconstruction algorithm \cite{Figueiredo}, the object $T$ can be reconstructed by solving the following convex optimization program:
\begin{eqnarray}
T = \left| {t'} \right|;{\rm{ \ }}
{\rm{ which \ minimizes: }}{\rm{ \ }}
\frac{{\rm{1}}}{{\rm{2}}}\left\| {I_t(x_i)  -
h (x)\otimes\left| {t'(x)} \right|^2 }
\right\|_2^2  \nonumber\\+ \tau \left\| {\rm{\Psi}}\{{\left| {t'(x)}
\right|^2}\} \right\|_1,{\rm{ }}\forall _s  = 1 \cdots M.
\end{eqnarray}
where $\tau$ is a nonnegative parameter which is used to adjust the sparse weight in the image restoration process, $M$ is the number of single-pixel detectors at the imaging plane $x_t$ and ${\rm{\Psi}}$ denotes the transform operator to the sparse basis. $\left\| V \right\|_{2}$ and $\left\| V \right\|_{ 1 }$ represent the Euclidean norm and the $\ell_1$-norm of $V$, respectively.
\begin{figure}[htbp]
\centering
\includegraphics[width=8.5cm]{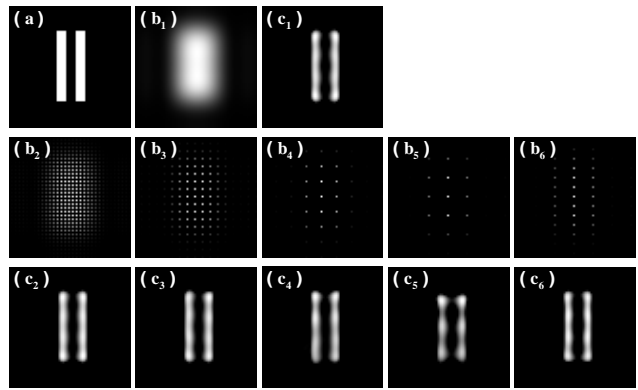}
\caption{The simulated demonstration of noiseless sub-Rayleigh imaging with sparse detection. (a) The original object; (b$_1$) the intensity distribution recorded by $M$=4096 sparse-array single-pixel detectors at the imaging plane $x_t$ and the distance between two singe-pixel detectors in lateral direction $\Delta x$=6.45 $\mu$m; (b$_2$) $M$=1024 and $\Delta x$=12.90 $\mu$m; (b$_3$) $M$=256 and $\Delta x$=25.80 $\mu$m; (b$_4$) $M$=64 and $\Delta x$=51.60 $\mu$m; (b$_5$) $M$=36 and $\Delta x$=64.50 $\mu$m; (b$_6$) $M$=72 and $\Delta x$=64.50 $\mu$m; (c$_1$-c$_6$) are corresponding sparse reconstruction results with respect to (b$_1$-b$_6$).}
\end{figure}
\begin{figure}[htbp]
\centering
\includegraphics[width=8.5cm]{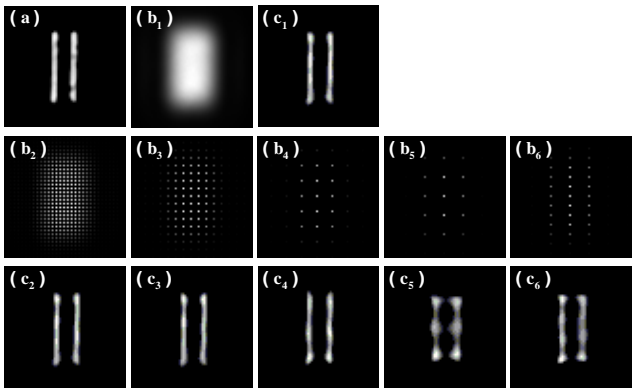}
\caption{The experimental demonstration of sub-Rayleigh imaging with sparse detection. (a) The original object imaged by a conventional optical imaging setup with large \emph{N.A.}; (b$_1$-b$_6$) and (c$_1$-c$_6$) are the intensity distributions record by the detectors at the imaging plane $x_t$ and their corresponding sparse reconstruction results, the same as Fig. 2.}
\end{figure}
\begin{figure}[htbp]
\centering
\includegraphics[width=8.5cm]{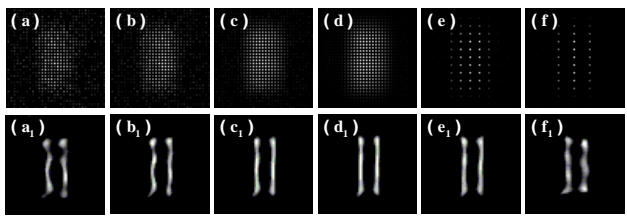}
\caption{Sub-Rayleigh imaging with sparse detection in different detection SNR and ($M$, $\Delta x$). (a) The intensity distributions recorded by $M$=1024 ($\Delta x$=12.90 $\mu$m) sparse-array single-pixel detectors and the detection SNR=2 dB at the imaging plane $x_t$; (b) $M$=1024 ($\Delta x$=12.90 $\mu$m) and SNR=5 dB; (c) $M$=1024 ($\Delta x$=12.90 $\mu$m) and SNR=10 dB; (d) $M$=1024 ($\Delta x$=12.90 $\mu$m) and SNR=15 dB; (e) $M$=120 ($\Delta x$=38.70 $\mu$m) and SNR=15 dB; (f) $M$=72 ($\Delta x$=64.50 $\mu$m) and SNR=15 dB; (a$_1$-f$_1$) are corresponding sparse restoration results with respect to (a-f).}
\end{figure}

In order to verify the idea, Fig. 2 and Fig. 3 have given the simulated and experimental demonstration of sub-Rayleigh imaging with sparse detection, using the optical system depicted in Fig. 1. The original object (64$\times$64 pixels, and the pixel size is 6.45 $\mu$m$\times$6.45 $\mu$m), as shown in Fig. 2(a) and Fig. 3(a), is a double-slit (slit width $a$=30 $\mu$m, silt height $h$=240 $\mu$m and center-to-center separation $d$=60 $\mu$m) and it comprises of 370 nonzero values in real-space domain. The parameters listed in Fig. 1 are set as follows: $z_1$=$z_2$=800 mm, the focal length of the lens $f$=400 mm and its effective transmission aperture $L$=8.0 mm. In addition, the pixel size of the single-pixel detector is 6.45 $\mu$m$\times$6.45 $\mu$m. Therefore, based on Eq. (1), the imaging system's resolution limit is $\Delta x_s \simeq$80 $\mu$m and the object's image at the imaging plane $x_t$ can not be resolved for conventional imaging [Fig. 2(b$_1$), Fig. 3(b$_1$)]. However, both the simulated and experimental results illustrated in Fig. 2 and Fig. 3 clearly demonstrate that sub-Rayleigh imaging with sparse detection can be achieved by exploiting the prior knowledge of the object's sparsity and the optical system's PSF. When the number of singe-pixel detectors (namely the total effective measurement number $M$) at the imaging plane $x_t$ is $M$=4096, 1024, 256, 64, 36, and 72, Fig. 2(b$_1$-b$_6$) and Fig. 3(b$_1$-b$_6$) present the intensity distributions recorded by the detectors, and their corresponding sparse reconstruction results borne on Eq. (4) are depicted in Fig. 2(c$_1$-c$_6$) and Fig. 3(c$_1$-c$_6$), respectively. As the distance between two singe-pixel detectors is increased, the restoration quality will be reduced because of the decrease of the sampling number [Fig. 2(c$_1$-c$_5$) and Fig. 3(c$_1$-c$_5$)]. However, if the sampling number at the imaging plane $x_t$ is increased, sub-Rayleigh imaging with sparse detection can be stably restored, even if the distance between two singe-pixel detectors ($\Delta x$=64.5 $\mu$m) is large than the double-slit's center-to-center separation [Fig. 2(c$_5$-c$_6$) and Fig. 3(c$_5$-c$_6$)]. In addition, as shown in Fig. 2(c$_6$) and Fig. 3(c$_6$), only $M$=72 measurements are used to restore the object's image, which is far fewer than suggested by Shannon's sampling theorem.

As shown in Fig. 4, we also perform the dependance of sub-Rayleigh imaging with sparse detection on the detection SNR at the imaging plane $x_t$. Fig. 4(a-f) present the intensity distributions recorded by the detectors at the imaging plane $x_t$ in different SNR. Based on the sparse reconstruction technique described by Eq. (4), the corresponding restoration results are displayed in Fig. 4(a$_1$-f$_1$). From Fig. 4(a$_1$-d$_1$), even if the detection SNR is 2 dB, we can approximately reconstruct sub-Rayleigh imaging using $M$=1024 measurements and the reconstruction quality will be improved as the increase of the detection SNR. While in the case of the detection SNR=15 dB, high-quality sub-Rayleigh imaging with sparse detection, as shown in Fig. 4(d$_1$-f$_1$), can be still restored by using the measurements far below Nyquist rate.

In conclusion, we have realized single-shot sub-Rayleigh imaging with sparse detection, exploiting the imaging object's sparsity and the prior knowledge of fixed optical system. Both the simulated and experimental results have demonstrated that using the measurement below Nyquist rate, sub-Rayleigh imaging with sparse detection can break through the limitation of both the optical system's Rayleigh limit and the camera's pixel-resolution to imaging resolution. Furthermore, the experiment has also shown that sub-Rayleigh imaging with sparse detection can still be approximately reconstructed in the case of measured SNR=2dB. This technique is very useful to microscopy of living cells or bacteria, and imaging of atoms captured by ion trap, etc, where the images are enough sparse.

The work was supported by the Hi-Tech Research and Development Program of China under Grant Project No. 2011AA120101 and No. 2011AA120102.

\end{document}